\documentstyle[preprint,aps]{revtex}
\tightenlines
\catcode`@=11
\def\references{%
\ifpreprintsty
\bigskip\bigskip
\hbox to\hsize{\hss\large \refname\hss}%
\else
\vskip24pt
\hrule width\hsize\relax
\vskip 1.6cm
\fi
\list{\@biblabel{\arabic{enumiv}}}%
{\labelwidth\WidestRefLabelThusFar  \labelsep4pt %
\leftmargin\labelwidth %
\advance\leftmargin\labelsep %
\ifdim\baselinestretch pt>1 pt %
\parsep  4pt\relax %
\else %
\parsep  0pt\relax %
\fi
\itemsep\parsep %
\usecounter{enumiv}%
\let\p@enumiv\@empty
\def\theenumiv{\arabic{enumiv}}%
}%
\let\newblock\relax %
\sloppy\clubpenalty4000\widowpenalty4000
\sfcode`\.=1000\relax
\ifpreprintsty\else\small\fi
}
\catcode`@=12

\begin{document}

\font\fortssbx=cmssbx10 scaled \magstep2
\hbox to \hsize{
%\special{psfile=uwlogo.ps hscale=8000 vscale=8000 hoffset=-12 voffset=-2}
%\hskip.5in \raise.1in
\hbox{\fortssbx University of Wisconsin - Madison}
\hfill\vtop{\hbox{\bf MADPH-97-1015}
            \hbox{\bf UCD-97-23}
            \hbox{October 1997}} }

\vspace*{.5in}

\begin{center}
{\large\bf  Relaxing Atomic Parity Violation Constraints on New Physics}\\
\vskip1cm
V. Barger$^1$, Kingman Cheung$^2$, D.P. Roy$^{3,4}$, and D. Zeppenfeld$^1$\\
\vskip.5cm
$^1${\it Department of Physics, University of Wisconsin, Madison, WI 53706}\\
$^2${\it Department of Physics, University of California, Davis, CA 95616}\\
$^3${\it Department of Physics, University of California, Riverside, CA
92521}\\
$^4${\it Tata Institute of Fundamental Research, Mumbai 400 005, India}
\end{center}

\vskip1in

\begin{abstract}
The weak charge $Q_W$ measured in atomic parity violation experiments can
receive compensating contributions from more than one new physics source.
We show explicitly that the $\Delta Q_W$ contribution from the exchange of an
extra $Z$-boson can cancel that from the $s$-channel scalar top or scalar
charm exchange in $R$-parity violating SUSY models proposed to explain the
HERA high-$Q^2$ anomaly.
\end{abstract}

\thispagestyle{empty}
\newpage

Parity violation in the Standard Model results from exchanges of weak
gauge bosons.  In electron-hadron neutral current (NC) processes parity
violation is
due to vector axial-vector ($VA$) and axial-vector vector ($AV$) interaction
terms in the Lagrangian. These interactions are tested
at the percent level at low momentum transfers ($Q^2\approx 0$) by the latest
atomic parity violation (APV)
measurements~\cite{apv} and at high momentum transfers
($Q^2\agt2,500$~GeV$^2$) by deep inelastic NC scattering at HERA. The
recently published NC data from the H1
experiment~\cite{hera} raise the possibility
of a scalar resonance in $e^+q\to e^+q$ scattering with mass
$M_{\tilde q} \approx200$~GeV~\cite{LQ}.

Given the high precision of the APV measurements, parity violating new
physics interpretations of the HERA high-$Q^2$ ``anomaly'' are fairly tightly
constrained.
A recent survey of the situation~\cite{altarelli} concludes that in $R$-parity
violating SUSY models an $s$-channel resonance
interpretation of the H1 events is only marginally consistent with APV
measurements.
In this brief note we examine this issue and point out that richer models
of new physics, which contain new particles beyond an $eq$ resonance,
can quite naturally relax the constraints from APV measurements. The two
extra contributions that we
consider are the exchange of an extra $Z$ boson and the exchange of squarks in
the crossed channel.

In low-momentum transfer NC processes, the $Z$ boson exchange is
well approximated by effective four-fermion contact terms. The parity
violating part of the NC interaction Lagrangian is conventionally
parametrized by constants $C_{1q}$ and $C_{2q}$ as
\begin{equation}
\label{lag}
{\cal L}^{e \,{\rm Hadron}} = \frac{G_F}{\sqrt{2}} \; \sum_q \biggr[
C_{1q} \left(\bar e \gamma^\mu \gamma^5 e\right) \; \left(\bar q \gamma_\mu
q\right) + C_{2q} \left(\bar e \gamma^\mu  e \right)\; \left(\bar q \gamma_\mu
\gamma^5 q\right) \biggr] \;.
\end{equation}
APV experiments are mostly sensitive to $C_{1q}$, for which the
radiatively corrected SM values are given by~\cite{pdg}
\begin{equation}
C_{1q}^{\rm SM} = \rho'_{eq}\left[-T_{3q} +2Q_q (\kappa'_{eq}
\sin^2\theta_{\rm w})\right]\; ,
\label{C1q}
\end{equation}
where $\sin^2\theta_{\rm w} = 0.2236$, $\rho'_{eq}=0.9884$,
and $\kappa'_{eq}=1.036$.

Atomic parity violation has been measured by several
methods~\cite{langacker}. The most  recent and precise experiment measures a
parity-odd atomic transition in Cesium atoms~\cite{apv}.   The advantage of
using the heavy Cs atom, with only a single valence electron,
is the smallness of the theoretical uncertainty due to atomic wave-function
effects.

APV experiments probe the weak charge $Q_W$ that parametrizes the parity
violating Hamiltonian~\cite{ham}
\begin{equation}
{\cal H}_{\rm APV} = {G_F\over2\sqrt 2} \, Q_W \, \rho_{\rm nucleus}({\bf r})
\gamma_5 \,.
\end{equation}
In terms of the parameters $C_{1u}$ and $C_{1d}$ of the NC Lagrangian
(\ref{lag}), the weak charge is given by~\cite{langacker}
\begin{equation}
Q_W = -2 \biggr[ C_{1u} (2Z+N) + C_{1d} (Z+2N) \biggr] \;,
\label{Q_W}
\end{equation}
where $Z$ and $N$ are the number of protons and neutrons in the nucleus
of the atom, respectively.
For $^{133}_{\phantom055}$Cs, the relation of $Q_W$ to the $C_{1q}$ is
\begin{equation}
Q_W = - 376 C_{1u} - 422 C_{1d} \;.
\end{equation}
With the radiatively corrected $C_{1q}$ of (\ref{C1q}),  the SM value of
$Q_W$ for Cs is~\cite{cho}
\begin{equation}
\label{sm}
Q_W^{\rm SM} = -73.11 \pm 0.05 \;.
\end{equation}
The recent precise measurement on Cesium atoms~\cite{apv} finds
\begin{equation}
\label{exp}
Q_W^{\rm exp} = -72.11\pm 0.27 \pm 0.89 \;,
\end{equation}
where the first error is statistical and the second one is theoretical.
This result is a substantial improvement from the value in the 1996
Particle Data Book \cite{pdg}
and shows better agreement with the SM than previously.
The $Q_W$ measurement places strong constraints on possible new physics
contributions~\cite{ours,dean},  $\Delta C_{1u}$ and
$\Delta C_{1d}$, that give
\begin{equation}
\label{delta}
\Delta Q_W \equiv Q_W - Q_W^{\rm SM} = -2 \biggr[ \Delta C_{1u} (2Z+N) + \Delta
C_{1d}(Z+2N)  \biggr] \;.
\end{equation}
{}From (\ref{sm}) and (\ref{exp}) one obtains
\begin{equation}
\Delta Q_W = 1.00\pm0.93 \,,  \label{dqw}
\end{equation}
where the stated uncertainty combines the statistical and theoretical errors in
quadrature. The central value of $\Delta Q_W$ is about $1\sigma$ above zero.

\section*{Squarks with {\boldmath $R$}-parity violating couplings}

The H1 and ZEUS \cite{hera}
experiments at HERA observed an excess of events above SM
expectations at high momentum transfer squared ($Q^2 > 15{,}000\rm~GeV^2$).
Although the excess is only at a $2\sigma$ statistical level, this potential
anomaly has stimulated a large number of new physics interpretations
that have focused mainly on an $s$-channel exchange of a squark
in supersymmetry with $R$-parity violating couplings \cite{LQ} and on contact
interactions representing particle exchanges of mass-squared much larger than
$Q^2$ \cite{contact}. A recent comprehensive fit~\cite{ours}
of all low and high energy data relevant
to $eeqq$ contact interactions found that contact terms can improve
the description of the HERA data. However, once the most recent Drell-Yan
data~\cite{DY} from the Tevatron are considered as well, contact terms do not
improve the overall quality of the fit compared to the SM~\cite{ours}. Thus,
$s$-channel squark exchange remains the most attractive interpretation of
the HERA events if the anomaly exists. The $s$-channel production of a
squark of mass $M_{\tilde q} \approx 200$~GeV could account for the
excess events in the $187.5<M<212.5$~GeV mass region seen by H1
(8~events observed, 1.5~events expected), but not by ZEUS (3~events
observed, 3~events expected) \cite{hera}.

The squark interpretation faces severe constraints from direct searches
for first generation leptoquarks
at the Tevatron~\cite{d0,cdf} and from the APV measurement~\cite{apv}.
The CDF and D0 experiments rule out squarks of mass up to
213 and 225 GeV, respectively, at 95\% CL, that decay
with branching fraction $B=100$\% into $e q$.
In order for a squark with $M_{\tilde q}\approx200$~GeV
to be consistent with the Tevatron limits,
the branching fraction is bounded from above by~\cite{d0,cdf}
\begin{equation}
\label{up}
B\alt 0.6 \;.
\end{equation}
The APV measurement, on the other hand, puts a lower limit on $B$, which we
will consider shortly.

The relevant term in the superpotential for the $R$-parity violating squark
explanation of the HERA anomaly is $\lambda'_{ijk} L_i Q_j
\overline{D_k}$.  The corresponding terms in the Lagrangian are
\begin{eqnarray}
{\cal L}_{ L_i Q_j \overline{D_k}} & =& \lambda'_{ijk} \biggr[
\tilde{e}_{iL} \overline{d_{kR}} u_{jL} +
\tilde{u}_{jL} \overline{d_{kR}} e_{iL} +
\tilde{d^*}_{kR} \overline{(e_{iL})^c} u_{jL} \nonumber \\
& & -
\tilde{\nu}_{iL} \overline{d_{kR}} d_{jL}
-\tilde{d}_{jL} \overline{d_{kR}} \nu_{iL}
-\tilde{d^*}_{kR} \overline{(\nu_{iL})^c} d_{jL} \biggr] + h.c. \;
\label{eq:Lyukawa}
\end{eqnarray}
where $i,j,k$ are the family indices, and $c$ denotes the charge conjugate.
The effective Lagrangians for the $ed$ and $eu$ scattering in the low-energy
limit are
\begin{eqnarray}
{\cal L}_{ed} &=& \frac{{\lambda'_{ijk}}^2}{M_{\tilde{u}_{jL}}^2}
 \left( \overline{e_{iL}} d_{kR} \right )
 \left( \overline{d_{kR}} e_{iL} \right )  \\
{\cal L}_{eu} &=& \frac{{\lambda'_{ijk}}^2}{M_{\tilde{d}_{kR}}^2}
 \left( \overline{(e_{iL})^c} u_{jL} \right )
 \left( \overline{u_{jL}} (e_{iL})^c \right ) \;.
\end{eqnarray}
By making a Fierz transformation these terms can be cast into a product of
leptonic and hadronic vector- or axial-vector currents, as in (\ref{lag}).
The resulting squark contributions to $\Delta C_{1q}$ are given by
\begin{equation}
\Delta C_{1d} = {\sqrt{2}\over G_F}\;
\left( \frac{ {\lambda'_{1j1}}^2 }{8 M_{\tilde{u}_{jL}}^2 } \right )
\;, \qquad
\Delta C_{1u} = - {\sqrt{2}\over G_F}\;
\left( \frac{ {\lambda'_{11k}}^2 }{8 M_{\tilde{d}_{kR}}^2 } \right )
\;.
\end{equation}
%
%%
%\begin{equation}
%\Delta C_{1d} = M_Z^2 \, \cos^2 \theta_{\rm w} \, \sin^2\theta_{\rm w}\,
%\left( \frac{ {\lambda'_{1j1}}^2 }{e^2 M_{\tilde{u}_{jL}}^2 } \right )
%\;, \qquad
%\Delta C_{1u} = - M_Z^2 \, \cos^2 \theta_{\rm w} \, \sin^2\theta_{\rm w}\,
%\left( \frac{ {\lambda'_{11k}}^2 }{e^2 M_{\tilde{d}_{kR}}^2 } \right )
%\;,
%\end{equation}
%%
%where $e^2 = 4\pi \alpha_{\rm em}$.
which cause a shift in $\Delta Q_W$ of:\footnote{Note that $\Delta Q_W$ does
not constrain $e^+s\to\tilde t_L$ production \cite{altarelli},
which is another viable mechanism to explain the HERA anomaly.}
\begin{equation}
\label{123}
\Delta Q_W = (2.4\;{\rm TeV})^2 \; \left[
\frac{ {\lambda'_{11k}}^2 }{M_{\tilde{d}_{kR}}^2 }  -
1.12 \; \frac{  {\lambda'_{1j1}}^2 }{M_{\tilde{u}_{jL}}^2 } \right]
\;.
\end{equation}

In order to account for the observed rate of the anomalous HERA high-$Q^2$
events with $e^+ d \to \tilde{t}_L / \tilde{c}_L$ production, the coupling
must be \cite{altarelli}
\begin{equation}
\label{pp}
\lambda'_{131} \;\;{\rm or}\;\; \lambda'_{121} \simeq
{0.03\over\sqrt{B}}
\end{equation}
for which (\ref{123}) gives
\begin{equation}
\Delta Q_W \approx -{0.14\over B} \,.  \label{dqw2}
\end{equation}
At the $2\sigma$ level the  APV measurement requires $\Delta Q_W > -0.87$
(implying  $\lambda'_{131}$ or $\lambda'_{121} < 0.074$ for $M_{\tilde q}\simeq
200$~GeV), bounding the $eq$ branching fraction from below by
\begin{equation}
\label{low}
0.2 \alt B \;.
\end{equation}
Combining the constraints in Eqs.~(\ref{up}) and (\ref{low}), $B$ is
restricted to the range
\begin{equation}
0.2 \alt B \alt 0.6 \;.
\end{equation}

In Ref.~\cite{roy} it was pointed out that most of the parameter
space of the minimal supersymmetric standard model (MSSM), with universal
masses at the unification scale, gives $B$ of order
0.1.  With $B\simeq0.1$ the constraint from the APV measurement
would be violated at the $3\sigma$ level.
Since up- and down-type squarks contribute with opposite sign  to
$\Delta Q_W$ [see (\ref{123})], can the $\Delta Q_W$ conflict be resolved by a
cancellation of the squark contributions?

The answer is yes, but marginally so. According to (\ref{eq:Lyukawa})
a $\tilde{d}_{kR}$ couples to both $e^-_Lu_L$ and $\nu_Ld_L$ and thus
$\tilde{d}_{kR}$ exchange contributes to CC observables. One finds that
$\lambda'_{111}$ is constrained to be less than 0.00035 from double-beta
decay~\cite{doublebeta,dreiner} (for a squark mass of 100~GeV) and thus is
irrelevant to our considerations. The
$\lambda'_{112,113}$ are constrained by charged-current universality
to~\cite{dreiner,bgh}
\begin{equation}
|\lambda'_{112,113}| < 0.02 {M_{\tilde{d}_{kR}}\over 100 {\rm GeV}}\; .
\label{eq:lambdalim}
\end{equation}
With (\ref{123}) the maximal contribution of a $\tilde{d}_{kR}$ to $\Delta Q_W$
is
\begin{equation}
\Delta Q_W \approx +0.23
\end{equation}
and may thus cancel the contribution from an up-type squark in (\ref{dqw2}),
but only for large branching ratios $B$. Given the stringent constraint
on the $R$-parity violating couplings in (\ref{eq:lambdalim}) it is unlikely
that a $\tilde{d}_{kR}$ would have been observed in direct production in
$e^-p$ collisions at HERA for which each of the HERA experiments has
collected $\sim1\rm~pb^{-1}$ of data~\cite{lqlim}.

\section*{Extra {\boldmath $Z$} models}

The Lagrangian describing the SM $Z$ boson ($Z_1^0$) and an extra $Z$ boson
($Z_2^0$) can be written as~\cite{ourZ'}
\begin{equation}
\label{ZZ} -{\cal L}_{Z^0_1 Z^0_2} = g_1 Z^0_{1\mu}  \sum_i
 \bar \psi_i \gamma^\mu (g_L^{i(1)} P_L + g_R^{i(1)} P_R) \psi_i +
 g_2 Z^0_{2\mu}   \sum_i
 \bar \psi_i \gamma^\mu (g_L^{i(2)} P_L + g_R^{i(2)} P_R) \psi_i \;,
\end{equation}
where $P_{L/R}= (1\mp \gamma_5)/2$, $g_1=e/(\sin\theta_{\rm w} \cos\theta_{\rm
w})$, $g^{i(1)}_L = T_{3i}-\sin^2 \theta_{\rm w} Q_i$ and
$g^{i(1)}_R=-\sin^2\theta_{\rm w} Q_i$,
$g_2/g_1=\sqrt{5 \sin^2\theta_{\rm w} \lambda/3}$ and $\lambda\simeq 1$.
In general, the SM $Z$ boson and the extra $Z$ boson will mix to form the
physical mass eigenstates $Z_1$ and $Z_2$,
\begin{equation}
\label{mixing}
\left ( \begin{array}{c} Z_1 \\
                         Z_2
        \end{array} \right ) = \left( \begin{array}{cc}
                                  \cos\theta & \sin\theta \\
                                 -\sin\theta & \cos\theta
                                      \end{array} \right ) \;
    \left( \begin{array}{c} Z^0_1 \\
                            Z^0_2
            \end{array} \right ) \;
\end{equation}
Here  $\theta$ is the mixing angle, $M_{Z_1}=91.1863$ GeV is the
mass of the $Z$ boson observed at LEP and SLC.
For simplicity we neglect the mixing since it is constrained to be small by
the LEP and SLC data at the $Z$ pole~\cite{ourZ'}.
In the zero mixing angle limit
the Lagrangian in Eq.~(\ref{ZZ}) describes
the interactions of physical $Z_1$ and $Z_2$ bosons.
%% equation (22) deleted
The contributions from the extra $Z$ boson to the coefficients $C_{1q}$ and
$C_{2q}$ are
\begin{equation}
\label{z-c1q}
\Delta C_{1q} = 2 \left(\frac{M_{Z_1}}{M_{Z_2}}\right)^2\left(\frac{g_2}{g_1}
\right)^2 \, g_a^{e(2)} g_v^{q(2)} \;, \qquad
\Delta C_{2q} = 2 \left(\frac{M_{Z_1}}{M_{Z_2}}\right)^2\left(\frac{g_2}{g_1}
\right)^2 \, g_v^{e(2)} g_a^{q(2)} \;,
\end{equation}
where $g_v =g_L+g_R$ and $g_a =g_L-g_R$. From these expressions we can
calculate $\Delta Q_W$ in terms of the mass $M_{Z_2}$  and the  couplings
$g_{L,R}^{f(2)}$ of the extra $Z$ boson.
Weakly-coupled extended gauge models, like $E_6$, give the coupling constant
$g_2$ on the order of the weak coupling constant $g_1 = e/\sin \theta_{\rm w}$.
We shall take $\lambda=1$ for which $g_2/g_1\simeq 0.62$.

\section*{Compensating contributions to {\boldmath$\Delta Q_W$}}

A low energy supersymmetry and an extra $Z$ boson with mass of order 1~TeV are
both natural consequences of string theory~\cite{L-S}. Then with $R$-parity
violating interactions both squark and
$Z_2$ exchanges would contribute to $\Delta Q_W$.
Their combined effect  on $\Delta Q_W$ is
\begin{equation}
\Delta Q_W =  (2.4\;{\rm TeV})^2 \Biggr [
\frac{ {\lambda'_{11k}}^2 }{M_{\tilde{d}_{kR}}^2 }  -
1.12 \; \frac{  {\lambda'_{1j1}}^2 }{M_{\tilde{u}_{jL}}^2 }
- 0.42\, \frac{g_a^{e(2)} }{ M_{Z_2}^2} \left( g_v^{u(2)} +
1.12  g_v^{d(2)} \right ) \Biggr ] \;.
\end{equation}
We can see  that the $Z_2$ contribution can make the overall
$\Delta Q_W$ positive.
For example, for a 1 TeV $Z_2$ with $g_a^{e(2)}=-1=-g_v^{u(2)}$
and $g_v^{d(2)}=0$, the $Z_2$ contribution to $\Delta Q_W$ is $+2.4$.
Then with $B(\tilde t_L\to e^+ d) = 0.1$ in (\ref{dqw2}) the combined
$\Delta Q_W$
contribution from $\tilde t_L$ and $Z_2$ is $\Delta Q_W = +1.0$, which is
the central value of the experimental measurement~(\ref{dqw}).

\section*{Summary}

We briefly summarize our main points.

(i) The deviation $\Delta Q_W$ of the cesium APV measurement from the SM is
positive, but the deviation is only $1\sigma$.

(ii) The $\Delta Q_W$ contribution of the scalar top or scalar charm via
$R$-parity violating $\tilde t_L e^+d $ or $\tilde c_L e^+d $
couplings are negative.

(iii) The $\Delta Q_W$ contributions of the scalar bottom or scalar strange
are positive, but they are likely too small to cancel the contribution from the
scalar top or scalar charm because of the tight constraints on their
couplings and masses.

(iv) Extra $Z$ boson contributions to $\Delta Q_W$ can naturally be positive
and sufficiently large to compensate negative contributions of scalar top or
scalar charm and make the overall $\Delta Q_W$ positive.

(v) In particular, a scalar top interpretation of the HERA anomaly with the
MSSM branching fraction of $B(\tilde t_L\to e^+ d)\approx 0.1$ is not excluded,
since positive extra $Z$ contributions to $\Delta Q_W$ may compensate the
negative contributions from the scalar top.

(vi) Our discussion applies similarly to leptoquark models for the HERA
anomaly~\cite{general}.

\section*{Acknowledgments}

D.P.~Roy thanks Fermilab for hospitality while this work was in progress. This
research was supported in part by the U.S.~Department of Energy under Grant
Nos.~DE-FG02-95ER40896 and DE-FG03-91ER40674, and in part by the University
of Wisconsin Research Committee with funds granted by the Wisconsin Alumni
Research Foundation.

\end{document}